\documentclass[11pt,A4paper]{revtex4-1}
\usepackage{amsmath,amsfonts,amssymb}
\usepackage{graphicx}
\usepackage{algorithm}
\usepackage{algorithmic}

\newcommand{\arctanh}[1]{\mathrm{arctanh}#1}

\renewcommand{\b}{\beta}

\newcommand{\s}{\sigma}

\newcommand{\comment}[1]{}
\newcommand{\Tr}{{\rm Tr}}

\newcommand{\ud}{\mathrm{d}}

\begin{document}
\title{Replica Cluster Variational Method: the Replica Symmetric
  solution for the 2D random bond Ising model}

\author{Alejandro Lage-Castellanos}
\affiliation{''Henri-Poincar\'e-Group'' of Complex Systems and
  Department of Theoretical Physics, Physics Faculty, University of
  Havana, La Habana, CP 10400, Cuba. and }

\author{Roberto Mulet} \affiliation{''Henri-Poincar\'e-Group'' of
  Complex Systems and Department of Theoretical Physics, Physics
  Faculty, University of Havana, La Habana, CP 10400, Cuba}

\author{Federico Ricci-Tersenghi} \affiliation{Dipartimento di Fisica,
  INFN -- Sezione di Roma 1 and CNR -- IPCF, UOS di
  Roma,\\ Universit\`{a} La Sapienza, P.le A. Moro 5, 00185 Roma,
  Italy}

\author{Tommaso Rizzo} \affiliation{CNR -- IPCF, UOS di Roma,
  Universit\`{a} La Sapienza, P.le A. Moro 5, 00185 Roma, Italy}

\date{\today}

\begin{abstract}
  We present and solve the Replica Symmetric equations in the context
  of the Replica Cluster Variational Method for the 2D random bond
  Ising model (including the 2D Edwards-Anderson spin glass model).
  First we solve a linearized version of these equations to obtain the
  phase diagrams of the model on the square and triangular
  lattices. In both cases the spin-glass transition temperatures and
  the tricritical point estimations improve largely over the Bethe
  predictions. Moreover, we show that this phase diagram is consistent
  with the behavior of inference algorithms on single instances of the
  problem.  Finally, we present a method to consistently find
  approximate solutions to the equations in the glassy phase. The
  method is applied to the triangular lattice down to $T=0$, also in
  the presence of an external field.
\end{abstract}

\maketitle

\section{Introduction}
\label{sec:Intro}

Since the celebrated work of Edwards and Anderson in 1975
\cite{EdwaAnde} many efforts have been devoted to the analytic
description of spin glasses. Very remarkable is the solution found by
Parisi in 1979 \cite{a1,ParRSB} to the Sherrington-Kirkpatrick
mean-field model \cite{SherKirk78}. The physical interpretation of the
Parisi solution \cite{MPV} gave a solid basis to concepts like Replica
Symmetry (RS) and spontaneous Replica Symmetry Breaking (RSB), that
became of standard use in the scientific community.  The solutions of
many models, not necessarily of mean field type, were interpreted
along these ideas (see e.g.\ the review in \cite{JSPreview} about spin
glasses on finite dimensional lattices).

In this context the last decade has been very exciting both from the
conceptual and from the practical point of view. First, M\'ezard and
Parisi \cite{MP1,MP2} were able to solve analytically the spin glass
model on a Bethe lattice (usually called the Viana-Bray model
\cite{VB}) with a Replica Symmetry Breaking ansatz. Within this RSB
ansatz, the solution is given in terms of populations of fields that
contain all the necessary information to describe the low temperature
phase of the model. The extension to other models was immediate
\cite{ScienceSAT,KS,Col} and the approach was fundamental to the
introduction of the Survey Propagation algorithm \cite{KS} that has
been successfully applied in the solution of many {\em single
  instances} optimization problems \cite{Col2,VC3}. Moreover it was
soon recognized that the well-known Belief Propagation (BP) algorithm
\cite{Pearl} corresponds to the Bethe approximation \cite{KABASAAD},
that is the replica symmetric solution on the Bethe lattice.

Unfortunately all the above analytical results concern mean-field models.
To go beyond the Bethe approximation, one should consider also loops in the
interaction network and this turns out to be a highly non trivial task
(see for example \cite{MR,bolos1,bolos2,MooijKappen07,GKC10,Zhou}).
Yedidia and coworkers \cite{YFW05} described how to generalize the
Cluster Variational Method (CVM) of Kikuchi \cite{Kikuchi} that allows to
derive a free-energy that improves the Bethe one by considering exactly the
contribution of small loops. The minimization of the CVM free energy can be
achieved by the use of a Generalized Belief Propagation (GBP) algorithm
\cite{YFW05}, but the solution found is always replica symmetric.

The idea of merging the CVM with the RSB ansatz was around for some
years, but it remained elusive. Probably because the simplest
comprehension of the RSB ansatz within the Bethe approximation is
based on a probabilistic cavity construction \cite{MP1}, which is hard
(or even impossible) to derive for a general CVM.  In a recent paper
\cite{tommaso_CVM} we proposed a formal solution to this problem.  The
idea was to apply the CVM to an already replicated free energy, and
then within the RSB ansatz to send the number of replicas to
zero. This formulation allowed us to derive a set of closed equations
for some local fields, that play the same role of the cavity fields in
the Bethe approximation.  Unfortunately these fields enter into the
equations in an implicit form and so standard population dynamic
algorithms can not be used for finding the solution.  In previous
works \cite{tommaso_CVM, GBP_GF}, using linear stability analysis, we
showed that these equations improve the Bethe approximation on the
location of the phase boundaries. However the solution of these
equations in the low temperature phase, and the interpretation of this
solution in terms of the performance of inference algorithms are still
important open problems.

The main goal of this work is to extend our previous results in these
two directions. On the one hand, using a stability analysis we study
the phase diagram in the $\rho$ (density of ferromagnetic couplings)
versus $T$ (temperature) plane for the Edwards-Anderson model on the
square and triangular lattices.  Moreover, we show that the
Generalized Belief Propagation algorithm (GBP) stops converging close
to the spin-glass temperature predicted by our approximation.  On the
other hand, we propose an approximated method to deal, at the RS
level, with the complex equations that arise in the formalism in the
low $T$ phase.

The rest of the work is organized as follows.  In the next section, we
rederive the equations already obtained in \cite{tommaso_CVM} but now
limiting its scope to the RS scenario in the average case.  In section
\ref{sec:ResultsPD} we present the phase diagram obtained by a
linearized version of these equations and in \ref{sec:ResultsIN} we
study the consequences of this phase diagram for the perfomance of
GBP. Section \ref{sec:ResultsLT} show the solution of a non-linear
approximation for the RS equations in the glassy phase. Finally, the
conclusions and possible extensions of our approach are outlined in
section \ref{sec:Conc}.

\section{The CVM Replica Symmetric solution}
\label{sec:RSS}

The Edwards-Anderson model is defined by the Hamiltonian $H =
-\sum_{(ij)} J_{ij} s_i s_j - h \sum_i s_i$, where the first sum is
over neighboring spins on a finite dimensional lattice, the couplings
$J_{ij}$ are quenched random variables and $h$ is the external
field. Although the equations we write are valid for generic
couplings, our results will be obtained for couplings drawn from the
distribution $P(J) = \rho\, \delta(J-1) + (1-\rho) \delta(J+1)$.

In a model with quenched disorder the free-energy of typical samples
can be obtained from the $n\to 0$ limit of the replicated free-energy
\begin{multline}
  \Phi(n) = -{1\over n \beta \, N}\ln \Tr \left\langle \exp\Big(\sum_{(ij)} \beta
  J_{ij} \sum_{a=1}^n s_i^a s_j^a + \sum_i \beta h \sum_{a=1}^n s_i^a \Big)\right\rangle_J =\\
  = -{1\over n\beta \, N}\ln \Tr
  \exp\left(\sum_{(ij)}\ln \Big\langle \exp \beta J \sum_a s_i^a s_j^a
    \Big\rangle_J +\sum_i \b h \sum_{a=1}^n s_i^a \right)\;,
\label{eq:Phi}
\end{multline}
where $n$ copies of a system of $N$ spins are considered at inverse
temperature $\beta$, and the average over the quenched disorder is
represented by the angular brackets.

The starting point of the Kikuchi's CVM approximation is to choose a
set of regions of the graph over which the model is
defined. Restricting only to link and node regions, the cluster
variation method recovers Bethe approximation. We will concentrate here on three kind of
regions: plaquettes (square or triangles, depending on the lattice),
links and nodes. Using the definition, $\psi_r(\s_r)\equiv\prod_{i,j
  \in r }\langle \exp \beta J \sum_a s_i^a s_j^a \rangle_J$ the energy
of region $r$ is:
\begin{equation}
E_r=-\ln \prod_{ij}\psi_{ij}(\s_i,\s_j)-\ln \prod_i \psi_i(\s_i)\;,
\label{eq:Eregions}
\end{equation}
where the products run over all links and nodes (in presence of a
field) contained in region $r$. Let us also define the belief
$b_r(\s_r)$ as an estimation of the marginal probability of the
configuration $\s_r$ according to the Gibbs measure. Then, within this
approximation, the Kikuchi's free energy takes the form:
\begin{equation}
F_{K}=\sum_{r \in R}c_r \left( \sum_{x_r}b_r E_r+\sum_{x_r}b_r \ln b_r \right)\;,
\label{eq:Fbeliefs}
\end{equation}
where the so-called Moebius coefficient $c_r$ is the over-counting
number of region $r$ \cite{YFW05}. In the case of the EA in the square
lattice, the biggest regions are the square plaquettes, and by
definition $c_P = 1$. Since each link region is contained in two
plaquettes, $c_L = 1 - 2 = -1$. Moreover, the spins regions are
contained in 4 plaquettes and 4 links and $c_S = 1 - 4 \cdot c_P - 4
\cdot c_L = 1$. Similarly for the triangular lattices $c_P = 1$, $c_L
= 1 -2 \cdot c_P = -1$ and $c_S= 1 - 6 \cdot c_P -6 \cdot c_L = 1$.

Now, the Kikuchi free energy has to be extremized with respect to the
beliefs $b_r(\s_r)$, subject to the constraint that they are
compatible upon marginalization. For example, $b_{(ij)}(\s_i,\s_j) =
\sum_{\s_k,\s_l} b_{(ijkl)}(\s_i,\s_j,\s_k,\s_l)$ and $b_i(\s_i) =
\sum_{\s_j} b_{(ij)}(\s_i,\s_j)$ for the square lattice. It is already
a standard procedure \cite{YFW05,pelizzola05} to show that under these
conditions the beliefs can be written as:
\begin{equation}
b_r(\s_r) \propto \psi_r(\s_r) \prod_{(r',s')\in M(r)} m_{r' s'}(\s_r)\;,
\label{eq:belief}
\end{equation}
where $M(r)$ is the set of connected pairs of regions $(r',s')$ such
that $r'\setminus s'$ is outside $r$ while $s'$ coincides either with
$r$ or with one of its subsets (descendants). For example, if $r$ is
one link in a square lattice, the product in (\ref{eq:belief})
contains the so-called messages $m$ from the two squares adjacent to
it, and the messages $m$ from the six other links connected to it
(three on each extreme). The messages $m_{rs}$ obey the following
equations:
\begin{equation}
m_{rs}(\s_s) \prod_{(r',s') \in M(r,s)} m_{r's'}(\s_s) \propto \sum_{\s_{r \setminus s}} \psi_{r\setminus s}(\s_r) \prod_{(r'',s'') \in M(r)\setminus M(s)} m_{r'' s''}(\s_r)\;,
\label{eq:yedmessage}
\end{equation}
where $M(r,s)$ is the set of connected pairs of regions $(r',s')$ such
that $r'$ is a descendant of $r$ and $s'$ is either region $s$ or a
descendant of $s$.

For the particular cases we are considering here (2D square and
triangular lattices) the general expression (\ref{eq:yedmessage})
translates into the following two couple equations. The first equation
is identical for both lattices and reads
\begin{equation}
m_{(ij)\rightarrow j}(\s_j) \propto \sum_{\s_i} \psi_{(ij)}(\s_i,\s_j) M_{\alpha \rightarrow (ij)}(\s_i,\s_j)  M_{\beta \rightarrow (ij)}(\s_i,\s_j) \prod_{k \in \partial i \setminus j} m_{(ki)\rightarrow i}(\s_i)\;,
\label{eq:yed1}
\end{equation}
where $\alpha$ and $\beta$ are the two plaquette sharing the link
$(ij)$ and $\partial i$ is the set of neighbors of site $i$. The
notation used in this equation should make clear that messages are
sent between a region and one of its descendant.  The second equation
takes slightly different forms for the square and triangular lattices,
and we write it explicitly for the triangular lattice:
\begin{multline}
M_{(ijk) \to (ij)}(\s_i,\s_j) m_{(ik)\to i}(\s_i) m_{(jk)\to j}(\s_j) \propto
\sum_{\s_k} \psi_{(ik)}(\s_i,\s_k) \psi_{(jk)}(\s_j,\s_k)\\
\prod_{\alpha \in \partial(ik) \setminus (ijk)} M_{\alpha \to (ik)}(\s_i,\s_k) \prod_{\beta \in \partial(jk) \setminus (ijk)} M_{\beta \to (jk)}(\s_j,\s_k) \prod_{l \in \partial k \setminus \{i,j\}} m_{l \to k}(\s_k)\;,\label{eq:yed2}
\end{multline}
where, in practice, the first two products only contain one message
each. For the square lattice the equation modifies slightly and
contains some more products; disregarding all indices and arguments,
its schematic form is $M\,m\,m \propto \sum \psi\,\psi\,\psi\prod
M\prod M\prod M\prod m\prod m$.

Up to this point the only difference with the standard CVM method is
the introduction of replicated spins $\s_i$ and the non obvious
connection with the average over the disorder, implicitly introduced
in $\psi_r(\s_r)$. The main contribution of our previous work
\cite{tommaso_CVM} was to introduce a consistent scheme to write these
equations in the limit $n \rightarrow 0$ at any level of RSB.

Here we reproduce the approach for the average case at the RS level. Following \cite{Mon98}, we start by parametrizing the link to node messages in the following way:
\begin{equation}
m(\s_i) = \int du\,q(u) \exp\left[\beta u\sum_{a=1}^n\sigma_i^a\right] (2\cosh \beta u)^{-n}\;,
\label{eq:parames1}
\end{equation}
and extend the same idea to the parametrization of the plaquette to link messages:
\begin{equation}
M(\s_i,\s_j) \propto \int dU\,du_i\,du_j\,Q(U,u_i,u_j) \exp\left[\beta U\sum_{a=1}^n\sigma_i^a\sigma_j^a+ \beta u_i\sum_{a=1}^n\sigma_i^a+ \beta u_j\sum_{a=1}^n\sigma_j^a\right]\;.
\label{eq:parames2}
\end{equation}
The above parametrization allows to rewrite the message passing equations (\ref{eq:yedmessage}) in terms of $q(u)$ and $Q(U,u_1,u_2)$. Substituting equations (\ref{eq:parames1}) and (\ref{eq:parames2}) into (\ref{eq:yed1}) and (\ref{eq:yed2}) and sending $n \rightarrow 0$, we obtain, after some standard algebra,
\begin{eqnarray}
q(u) &=& \int \prod_i^k dq_i \prod_\alpha^p dQ_\alpha\,\langle\delta(u-\hat{u}(\#))\rangle_J\;,\nonumber\\
R(U,u_a,u_b) & \equiv & \int du_i\,du_j\,Q(U,u_i,u_j) q(u_a-u_i) q(u_b-u_j) = \label{eq:RSmess}\\
& = & \int \prod_i^K dq_i \prod_\alpha^P dQ_\alpha\,
\langle\delta(U-\hat{U}(\#)) \delta(u_a-\hat{u}_a(\#)) \delta(u_b-\hat{u}_b(\#))\rangle_J\;,\nonumber
\end{eqnarray}
where $k$ ($p$) and $K$ ($P$) correspond to the number of small $m$ (large $M$) messages that enter into each equation. The specific expressions for $\hat{u}(\#),\hat{U}(\#),\hat{u}_a(\#),\hat{u}_b(\#)$ depend on the lattice. The expressions for the triangular lattices are given in the next section and we refer the reader to reference \cite{dual} for similar formulas for the square lattice.

The next step is to solve the self-consistency equations in (\ref{eq:RSmess}). Then, once $q$ and $Q$ are known, the thermodynamical observables are well defined in term of these objects \cite{tommaso_CVM}. Unfortunately, since in (\ref{eq:RSmess}) the functions $Q$ and $q$ are convoluted, this problem can not be straightforwardly approached using standard population dynamics algorithm. One possible approach is to deconvolve $R$ using Fourier techniques to extract $Q$. Unfortunately, this approach suffers from strong instability problems. To use any numerical Fourier transform, one must have $R$ and $Q$ in form of histograms. But since $Q$ is not necessarily positive defined \cite{tommaso_CVM} the sampling of the messages becomes hard and the numerical errors due to the discretization of $Q$ combine with the errors due to the Fourier inversion process making difficult the convergence at low temperatures. To bypass these numerical problems we choose to solve these equations approximately. We perturb them in terms of small parameters around the paramagnetic solution and keep track of the information about the first few moments of the distributions.

\section{Phase Diagram from the linearized equations}
\label{sec:ResultsPD}

Since the exact computation of $q(u)$ and $Q(U,u_1,u_2)$ is a daunting task, here we concentrate our attention on the calculation of their first two moments:
\begin{eqnarray}
 && m = \int q(u)\,u \,\ud u\;,\qquad a = \int q(u) \,u^2 \ud u \;,\qquad
 a_0(U) = \iint Q(U,u_1,u_2) \, \ud u_1\, \ud u_2 \;, \nonumber\\
 && M_{i}(U) = \iint Q(U,u_1,u_2) \,u_i \, \ud u_i \;, \qquad
 a_{ij}(U) = \iint Q(U,u_1,u_2) \,u_i\,u_j \,\ud u_1 \ud u_2\;,
\label{eq:moments}
\end{eqnarray}
where $i,j \in \{1,2\}$. With these definitions the moments are determined by
\begin{eqnarray}
m & =&  \int \prod_{i}^k dq_i \prod_{\alpha}^p dQ_{\alpha}  \langle \hat{u} \rangle_J \\ \nonumber
a & =& \int \prod_{i}^k dq_i \prod_{\alpha}^p dQ_{\alpha}  \langle \hat{u}^2 \rangle_J \\ \nonumber
M_1(U)& =&  \int du_1 du_2 R(U,u_1,u_2)  \langle u_1 \rangle_J - m\,a_0(U)\\ \nonumber
a_{11}(U)& =&  \int du_1 du_2 R(U,u_1,u_2)  \langle u_1^2 \rangle_J- 2 m M_1(U) - a\,a_0(U)\\ \nonumber
a_{12}(U)& =& \int du_1 du_2 R(U,u_1,u_2) \langle  u_1 u_2 \rangle_J - 2 m M_1(U) - m^2 a_0(U)
\label{eq:Maa}
\end{eqnarray}
However, keep in mind that $R(U,u_1,u_2)$ is still defined in terms of $q$ and $Q$, see (\ref{eq:RSmess}), and not directly in terms of the moments. Therefore, in order to compute the integrals in (\ref{eq:Maa}), one must introduce some ansatz over these distributions. It is then reasonable to start considering as correct the high temperature solution and to linearize the equations around this solution. At high temperatures and zero external field one may assume that the system is paramagnetic:
\begin{equation}
q(u) = \delta(u)\;, \qquad
Q(U,u_1,u_2) = a_0(U) \delta(u_1)\delta(u_2)
\label{eq:parpa}
\end{equation}

In what follow we show, first, the linearization of  $\hat{u}(\#)$ and  $\hat{u}_i(\#)$ for the triangular lattice. Then, as an example, the derivation of the expressions for $m$ and $a$ in (\ref{eq:Maa}) and leave for the Appendix the expressions for the moments of $Q$.  The algebra associated to the equations for the square lattice is more cumbersome, but is technically equivalent. The interested reader may look for the case when $m=0$ in references \cite{tommaso_CVM} and \cite{dual}.

\begin{figure}[htb]
\includegraphics[width=0.75\textwidth]{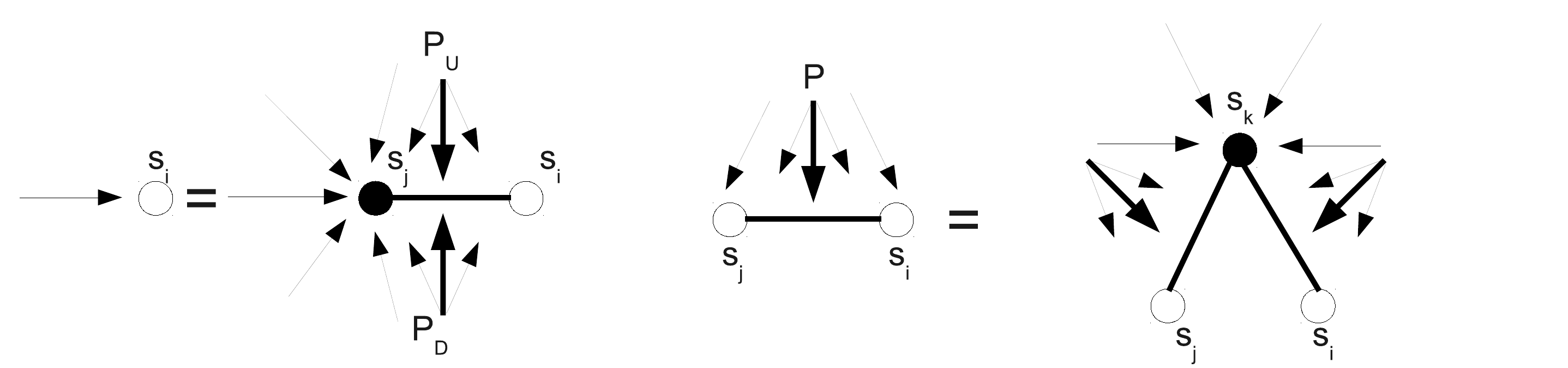}
\caption{Schematic representation of the message passing equations (\ref{eq:RSmess}) for the triangular lattice.}
\label{fig:Smess}
\end{figure}

To compute $\hat{u}(\#)$ it is enough to understand that the first equation in (\ref{eq:RSmess}) may be interpreted as a standard equation for the Bethe approximation with a renormalized interaction between the spins $\hat{J} = J + U^a+ U^b$ (see the first panel in Figure  \ref{fig:Smess}). Then, one can follow standard calculations \cite{MeMo} and the expression for the single message $\hat{u}$ reads
\begin{equation}
\hat{u}(\#) = \frac{1}{\beta}\arctanh[\tanh(\b \hat{J})\tanh(\b h)] + u_1^a+u_1^b\;,
\label{eq:u}
\end{equation}
where $u_1^a$ and $u_1^b$ are the small messages sent from the corresponding neighbor plaquettes to the site of interest and $h = u_2^a + u_2^b + \sum_i^5 u_i$  where $u_2^a$ and $u_2^b$ are the messages sent from the same plaquettes to the other border of the link. Considering that all the $u$'s and $h$ are small, as must be the case close to the paramagnetic transition, the linearized version of the previous expression becomes
\begin{equation}
\hat{u}(\#) = \tanh(\beta\hat{J}) h + u_1^a+u_1^b\;.
\label{eq:ulin}
\end{equation}

The messages in the second equation of (\ref{eq:RSmess}) can be rewritten through the following identities:
\begin{eqnarray}
 \hat{U}(\#) &=& \frac{1}{4 \b} \ln \frac{K(1,1) K(-1,-1)}{K(1,-1) K(-1,1)}\\\nonumber
 \hat{u}_1(\#) &=& \frac{1}{4 \b} \ln \frac{K(1,1) K(1,-1)}{K(-1,1) K(-1,-1)}\\\nonumber
 \hat{u}_2(\#) &=& \frac{1}{4 \b} \ln \frac{K(1,1) K(1,1)}{K(1,-1) K(-1,-1)}\\
\label{eq:Uuu}
\end{eqnarray}
where $K(S_1,S_2) = \sum_{S_3} \exp^{ \b \hat{J}_{13}S_1 S_3 + \b \hat{J}_{13}S_2 S_3 + u_1^a S_1 + u_1^b S_2 + h_3 S_3}$ (see the second panel in Figure \ref{fig:Smess}).
Then, after some algebra, it is easy to show that
\begin{eqnarray}
\hat{u}_1(\#)& =& \frac{1}{ 4 \beta}\ln\frac{(1+\tanh(\b J_+)\tanh(\beta h))(1+\tanh(\b J_{-})\tanh(\beta h))}{(1+\tanh(\b J_+)\tanh(\beta h))(1+\tanh(\b J_{-})\tanh(\beta h))} \\\nonumber
 &\sim &  u_1^a + \frac{1}{2} h [ \tanh(\beta J_{+}) + \tanh(\beta J_{-})]
\label{eq:ua}
\end{eqnarray}
and in a similar way
\begin{equation}
\hat{u}_2(\#) \sim u_1^b + \frac{1}{2} h [ \tanh(\beta J_{+}) - \tanh(\beta J_{-}) ]\;,
\label{eq:ublin}
\end{equation}
where $J_+ = (J^a + U^a) + (J^b+ U^b)$,  $J_{-} = (J^a + U^a) - (J^b + U^b)$ and $h=  u_2^a + u_2^b + \sum_i^3 u_i$.
With these expressions we have all the necessary ingredients to write the linearized form of (\ref{eq:Maa}). Next, we show how to derive the linear equations for $m$ and $a$ and in the Appendix we present the results for the others.

The single site magnetization $m= \langle u \rangle$ satisfies
\begin{multline}
m = \langle u \rangle = \int du q(u) u = \langle \int dQ_a dQ_b \prod_{i=1}^5 dq_i u \delta(u-\hat{u}(\#)) \rangle_J 
=\langle \int dQ_a dQ_b \prod_{i=1}^5 dq_i \hat{u} \rangle_J =\\
= \langle \int dQ_a dQ_b \prod_{i=1}^5 dq_i ( u_1^a+u_1^b+\tanh(\beta\hat{J}) h) \rangle_J
\label{eq:mlin}
\end{multline}
and using the definitions in (\ref{eq:moments}) the last integral can be easily expressed in linear terms of the moments of the distributions. The result is
\begin{multline}
m = 5 m \langle \int dU^a dU^b \tanh(\beta\hat{J})a_0(U^a)a_0(U^b) \rangle_J + \int dU^a M_1(U^a) + \int dU^b M_1(U^b) +\\
+\langle \int dU^a dU^b \tanh(\beta\hat{J}) [a_0(U^b) M_1(U^a) +  a_0(U^a)  M_1(U^b)]  \rangle_J
\label{eq:mlinF}
\end{multline}

The derivation of $a$ proceeds in a similar way
\begin{multline}
 a = \langle u^2 \rangle = \int du q(u) u^2 =  \langle \int dQ_a dQ_b \prod_{i=1}^5 dq_i ( u_1^a+u_1^b+\tanh(\beta\hat{J}) h)^2 \rangle_J = \\
\langle \int dQ_a dQ_b \prod_{i=1}^5 dq_i ( (u_1^a)^2+(u_1^b)^2+
2 u_1^a u_1^b + 2 ( u_1^a+u_1^b) \tanh(\beta\hat{J}) +  \tanh^2(\beta\hat{J}) h^2
 \rangle_J
\label{eq:alin}
\end{multline}
that may be re-written in term of the moments:
\begin{multline}
a = \int dU^a a_{11}(U^a) + \int dU^b a_{11}(U^b) + 2 m \int dU^a dU^b M_1(U^a) M_1(U^b) +\\
+10 m \langle \int dU^a dU^b \tanh(\beta\hat{J}) [ a_0(U^a) M_1(U^b)+ a_0(U^a) M_1(U^a) ]\rangle_J +\\
+2 \langle \int dU^a dU^b \tanh(\beta\hat{J})  [ a_{12}(U^a) + a_{12}(U^b)+ M_2(U^b) M_1(U^a) + M_1(U^b) M_2(U^a) ]\rangle_J+\\
+(5 a + 20 m^2) \langle \int dU^a dU^b \tanh^2(\beta\hat{J})a_0(U^a)a_0(U^b)\rangle_J+\\
+ 10 m \langle \int dU^a dU^b \tanh^2(\beta\hat{J}) [ a_0(U^a) M_2(U^b)+ a_0(U^a) M_2(U^a) ] \rangle_J +\\
+ \langle \int dU^a dU^b \tanh^2(\beta\hat{J}) [ a_{22}(U^a) + a_{22}(U^b)+ 2 M_2(U^b) M_2(U^a)] \rangle_J
\label{eq:alinF}
\end{multline}

Similar expressions may be derived for $M_i$ and $a_{ij}$, see Appendix, but note that they are not closed analytical expressions. The form of $a_0(U)$ is unknown, and must be determined for each $\beta$ using population dynamics. Once $a_0(U)$ has been computed, one can study the set of linear equations for the moments and check the local stability of the paramagnetic solution. In order to do this, we start from the paramagnetic solution, i.e. all the moments zero, but with $a_0(U)$ being non trivial. Then, we slightly perturb $a$ and $m$ and check, solving iteratively Eqs. (\ref{eq:mlinF}), (\ref{eq:alinF}) and (\ref{eq:M1trian})-(\ref{eq:a12trian}) whether these perturbations die out or diverge. Depending on $\rho$ and $T$ we find that under iteration, either both magnitudes diverge, or just $a$ or none. If $a$ and $m$ converge to zero the system is in the paramagnetic phase (P). If only $a$ diverges it is in the spin-glass phase (SG) and if both $a$ and $m$ diverge we say that the system is in a ferromagnetic phase (F). 

\begin{figure}[htb]
\includegraphics[angle=0,width=0.45\textwidth]{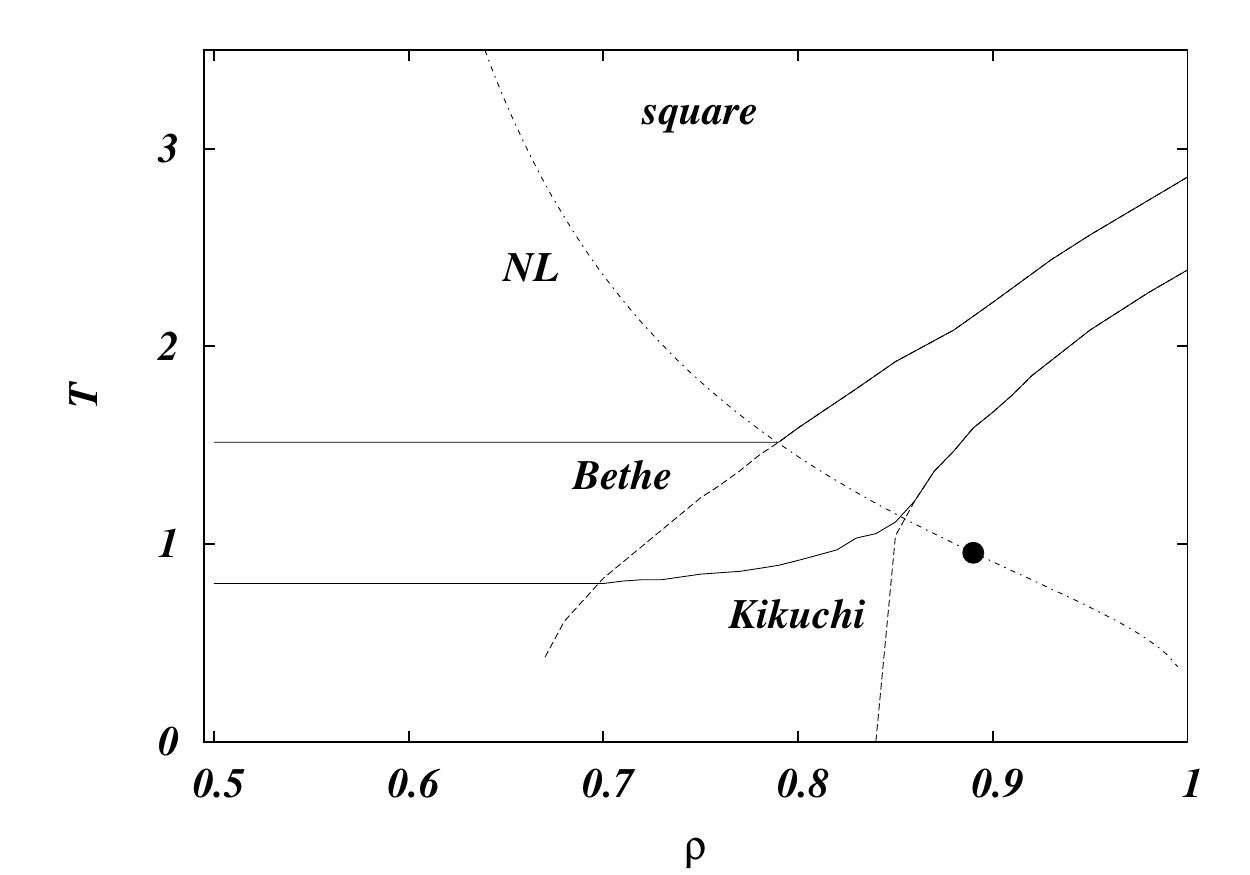}
\includegraphics[angle=0,width=0.45\textwidth]{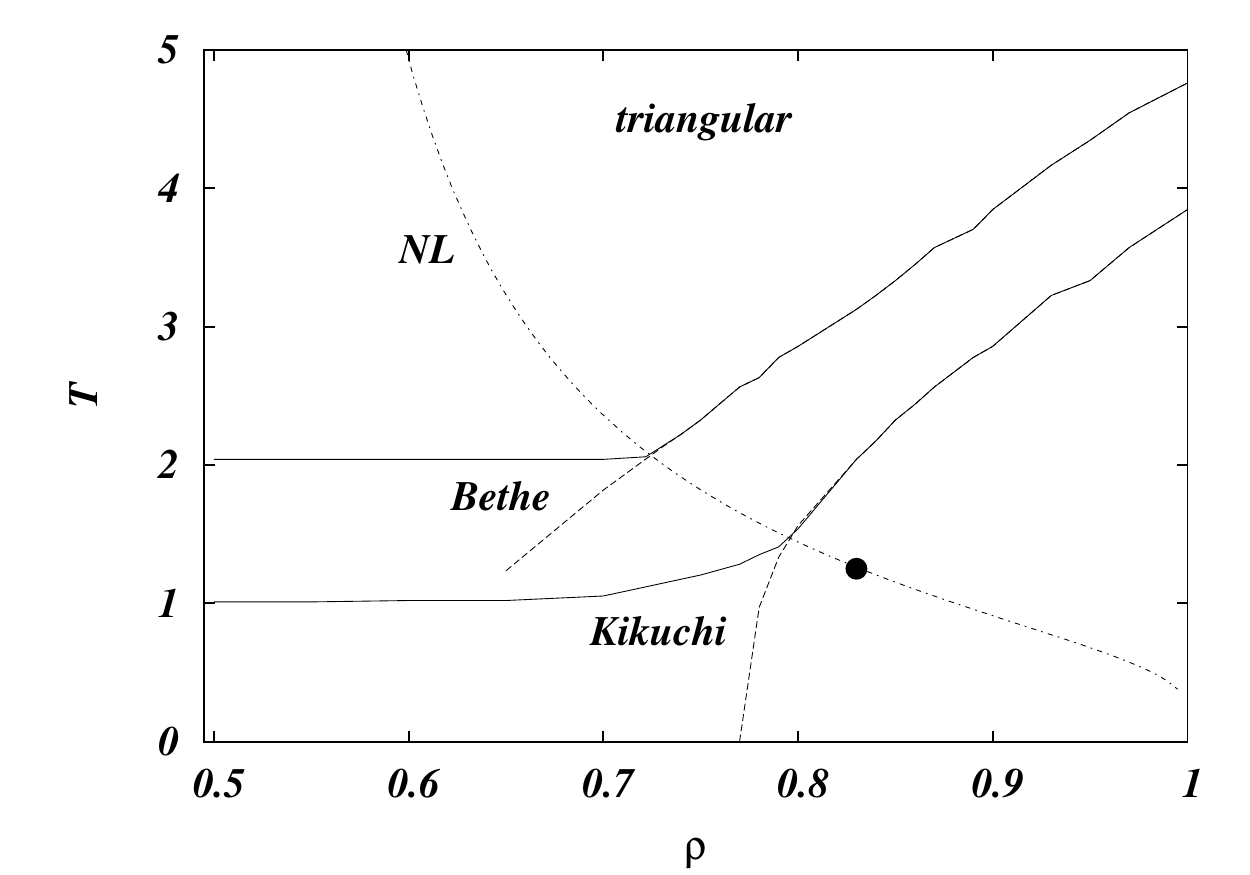}
\caption[0]{$\rho$ vs $T$ phase diagram for the square  and the triangular lattice. We show results for the Bethe approximation (upper curves) and the Kikuchi approximation (lower curves). It is also shown the Nishimori line (NL). The bold circles on the Nishimori line are the best analytical predictions for the tricritical points. The dashed lines represent the boundaries for the existence of purely ferromagnetic solutions.}
\label{fig:PhDiag}
\end{figure}

The results of this analysis are reported in Figure \ref{fig:PhDiag}. The phase diagrams must be read in the following way. Below the horizontal lines we have the Spin Glass phase and above the Paramagnetic phase. Critical lines meets at the tricritical point $(\rho^{cr},T^{cr})$, located on the Nishmori line (NL). On the right of this tricritical point, i.e.\ if $\rho > \rho_{cr}$, the system is in the Ferromagnetic phase at low temperatures and in the Paramagnetic phase at high temperatures.

In both cases, the conclusions are similar: the P-SG critical temperature predicted by the Kikuchi approximation is lower than the one predicted by the Bethe approximation. This result was already shown for $\rho=0.5$ in \cite{tommaso_CVM}, but here we correct an error in that work were an incomplete range of $\beta$ was considered during the study of the square lattice. In addition these results are now extended to larger values of $\rho$.  Moreover, we show that while both approximations correctly predict a SG to F transition at low temperatures and a tricritical point on the Nishimori line (NL), the estimation of the latter is much better in the Kikuchi approximation (the big dots on the NL are the exact locations for the tricritical points predicted in \cite{Nishi1} and \cite{Nishi2}). The following table summarize the locations of the tricritical points:
\begin{center}
\begin{tabular}{|l|c|c|r|} \hline
lattice & $\rho^{cr}_{Bethe}$ & $\rho^{cr}_{Kikuchi}$ & $\rho^{cr}_{exact}$\\ \hline
square & 0.79  & 0.85 & 0.8894 \\ \hline
triangular & 0.74 & 0.78 & 0.8358\\ \hline
\end{tabular}
\end{center}

Finally, we checked the existence of a ferromagnetic transition keeping $a$ zero and perturbing $m$. Again, Kikuchi approximation improves Bethe one. Indeed the latter predicts a SG-F critical line extending to very low $\rho$ values (well below $\rho_{cr}$), while the Kikuchi approximation have a SG-F critical line which is almost vertical in the $\rho$ vs $T$ phase diagram (and this behavior is consistent with the theoretical predictions \cite{Pfafian}).

\section{Connection to the behavior of inference algorithms}
\label{sec:ResultsIN}

The results so far presented are obtained by taking the average over the ensemble and should then correspond to properties of typical
samples in the large $N$ limit. It is known, however, that the predicted spin glass phase is not present in EA 2D at any finite temperature. This mistaken phase transition is a feature of any mean field like approximation (including Bethe and CVM), and therefore is not surprising. Nonetheless, the analytical method developed might keep its validity in relation to the behavior of message passing algorithms in single instances. In this section we explore this connection for models on the square lattice.

\begin{figure}[!htb]
\includegraphics[angle=270,width=0.48\textwidth]{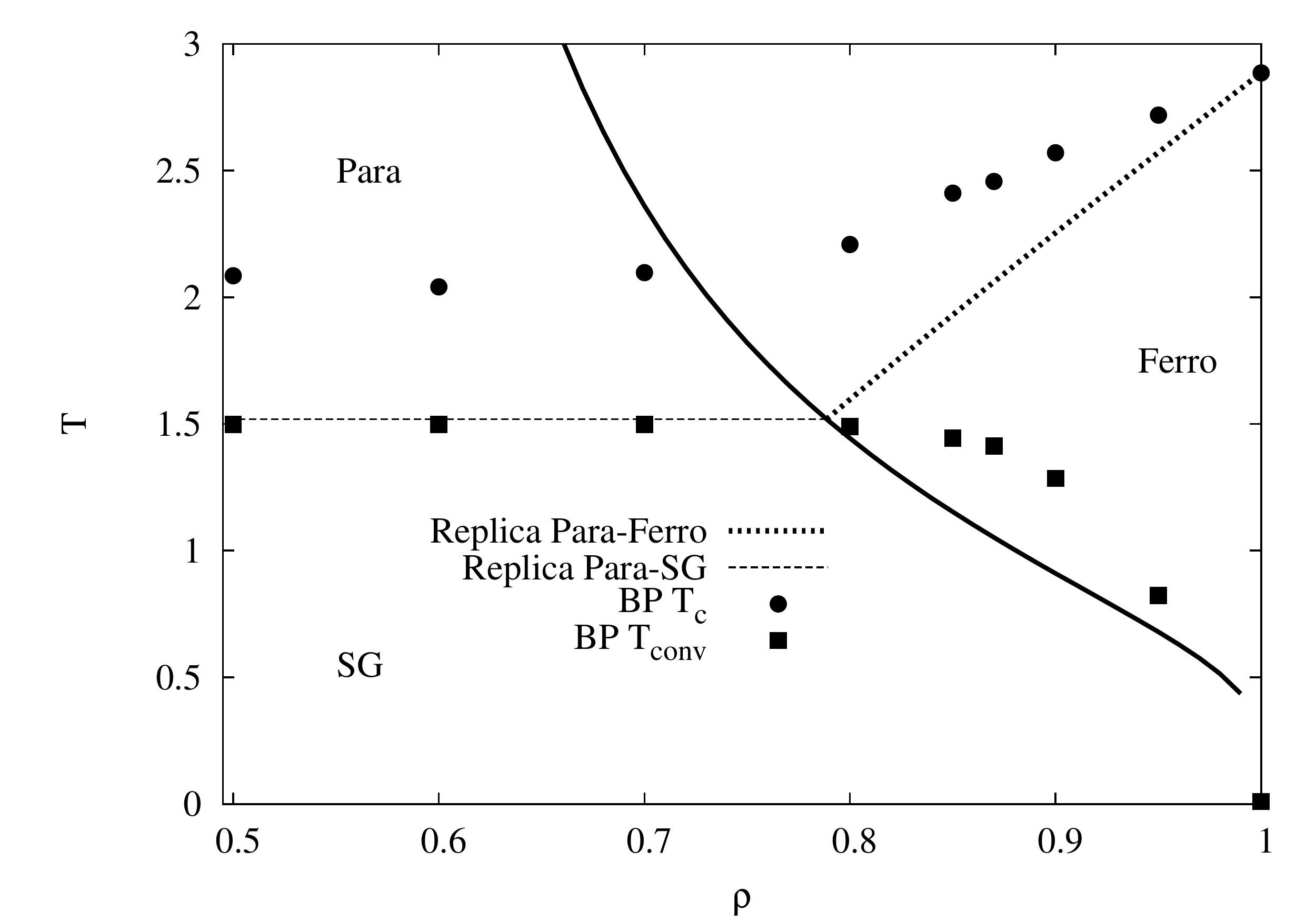}
\includegraphics[angle=270,width=0.48\textwidth]{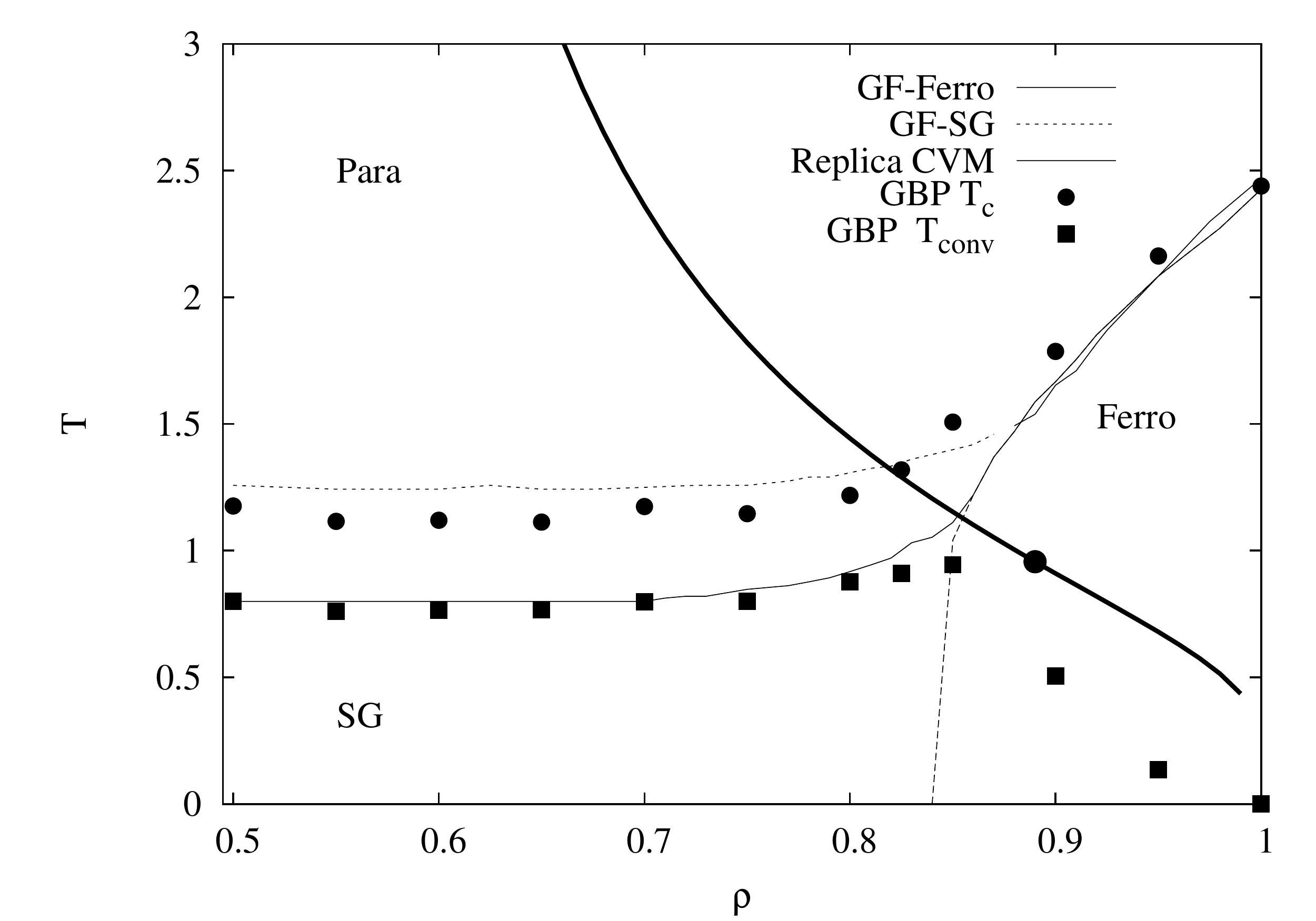}
\caption[0]{$\rho$ vs $T$ phase diagram for the square lattice in the Bethe (left panel) and plaquette-CVM (right panel) approximations. The circles indicate the temperature at which BP (GBP) finds a non paramagnetic solution and the squares the temperature below which
BP (GBP) does not converge. In the left panel 50 realizations of $N=64\times 64$ systems were used for the estimation of the critical temperatures, while in the right panel we have averaged over 10 samples of size $N = 256 \times 256$}
\label{fig:singleins}
\end{figure}

When running BP and GBP for the Bethe and plaquette-CVM approximations on the square lattice we find a paramagnetic solution at high temperatures, characterized by zero local magnetizations $m_i=0$. Below specific critical temperatures (that we call BP--$T_c$ and GBP--$T_c$) both algorithms find non paramagnetic solutions (i.e., with $m_i \neq 0$), as shown by the black circles in Fig. \ref{fig:singleins}. These critical temperatures in single instances are far from the values predicted by the replica method for the Para-SG and Para-Ferro transitions for $\rho<1$ values. As noticed in Ref.~\cite{zhouwang2012}, single instances of the Edwards-Anderson model present areas of low frustration where a disordered ferromagnetic state is found by BP and GBP. These regions are related to Griffith instabilities \cite{griffith69,vojta06} in finite dimensional disordered systems. It is, therefore, not surprising that the average case replica calculations, which are intrinsically homogeneous in space, fail to predict a transition related to this kind of singularities. It is worth noticing that below $T_c$ the solution found by BP has very small magnetizations (especially if compared with those found by GBP below GBP--$T_c$). This is the main reason why we missed BP--$T_c$ in Ref.~\cite{GBP_GF}.

\begin{figure}[htb]
\includegraphics[width=1.05\textwidth]{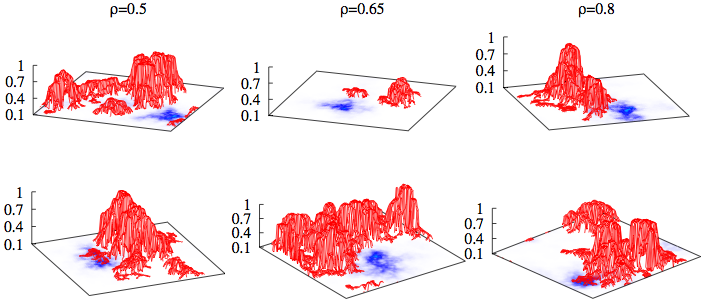}
\caption[0]{Red surfaces correspond to the absolute value of the
magnetization of the spins in a  $100\times 100$ system, while the
contours (shadowed blue areas) mark the regions where the non
convergence appears for the first time. For each of the three values of
$\rho$, two different realizations of the disorder are shown. The
xy-plane is set at magnetization $|m|=0.1$. In most cases the
convergence problems appear in the low magnetization regions.}
\label{fig:mag_stab}
\end{figure}

On the other hand, both  BP and GBP stop converging at a temperature that is quite close to the one predicted by the replica calculations for the Para-SG transition in the region $\rho<\rho^{cr}$ (see the black squares in Fig. \ref{fig:singleins}). Connecting the lack of convergence of an iterative algorithm (as GBP) to
the appearance of a flat direction in the CVM free-energy is something
very desirable: this is what one would call a `static' explanation to a
`dynamical' behavior. However here the situation is more subtle, because
on any given large sample the message passing algorithm (either BP or GBP) ceases to converge to the paramagnetic
fixed point at $T_c$: below $T_c$ the fixed point reached by BP and GBP has
many magnetized variables. So, how can the instability of the
paramagnetic fixed point (where all local magnetizations are null)
explain the lack of convergence of BP and GBP around the SG fixed point (with
non-null magnetization)? We have studied in detail the behavior of GBP
close to $T_\text{conv}$ and we have discovered that in the regions with
magnetized spins GBP messages are very stable and show no sign of
instability; on the contrary, in the regions where spin magnetizations
are very close to zero, the GBP messages start showing strong
fluctuations and finally produce an instability that leads to the lack
of convergence of GBP (see Fig. \ref{fig:mag_stab}). Since in these
regions of low local magnetizations the distribution of GBP messages is
very similar to the one of the paramagnetic fixed point, then the
average case computation for $T_\text{CVM}$ shown in the previous
Section may perfectly explain the divergence of GBP messages in these
regions. Again we have a `static' explanation for a `dynamical' effect,
and this is very desirable.

The above argument well explains the similarity between $T_\text{CVM}$
and $T_\text{conv}$ in the region $\rho < \rho^{cr}_{Kikuchi}$ where no
ferromagnetic long range order is expected to take place. However, for
$\rho > \rho^{cr}_{Kikuchi}$, the situation is more delicate: indeed
there is ferromagnetic long range order below the critical line, and so
the above argument can not hold as it is (there are no large regions
with null local magnetizations, where the instability can easily arise).
Moreover if we assume that a GBP
instability can mainly grow in a region of low magnetizations, we would
conclude that GBP must be much more stable for $\rho >
\rho^{cr}_{Kikuchi}$. Indeed what we see in Fig. \ref{fig:singleins} is
that the behavior of the filled squares drastically change around
$\rho^{cr}_{Kikuchi}$, and $T_\text{conv}$ becomes much smaller in the
ferromagnetic phase. This observation supports the idea that an
instability of GBP can mainly arise and grow in a region of low local
magnetizations: in a ferromagnetic phase these regions are rare and
small, and thus GBP is able to converge down to very low temperatures.

\subsection{Average case with population dynamics}

Recently in reference \cite{GBP_GF} we have studied in detail the
behavior of GBP on the 2D EA model (i.e. the present model with $\rho =
0.5$). For this particular case we reported the two important temperatures: a critical
temperature $T_c$ where the EA order parameter $q_{EA}$ predicted by the
GBP becomes different from zero and a lower temperature $T_\text{conv}$
where GBP stops converging to a fixed point. We noticed that the critical
temperature found by the replica CVM method was close to $T_
\text{conv}$, while a critical temperature close to $T_c$ could also be
obtained from an average case calculation based on a population dynamics
method, similar to the one used in \cite{MP1} for the Bethe approximation.

\begin{figure}[htb]
\includegraphics[angle=0,width=0.3\textwidth]{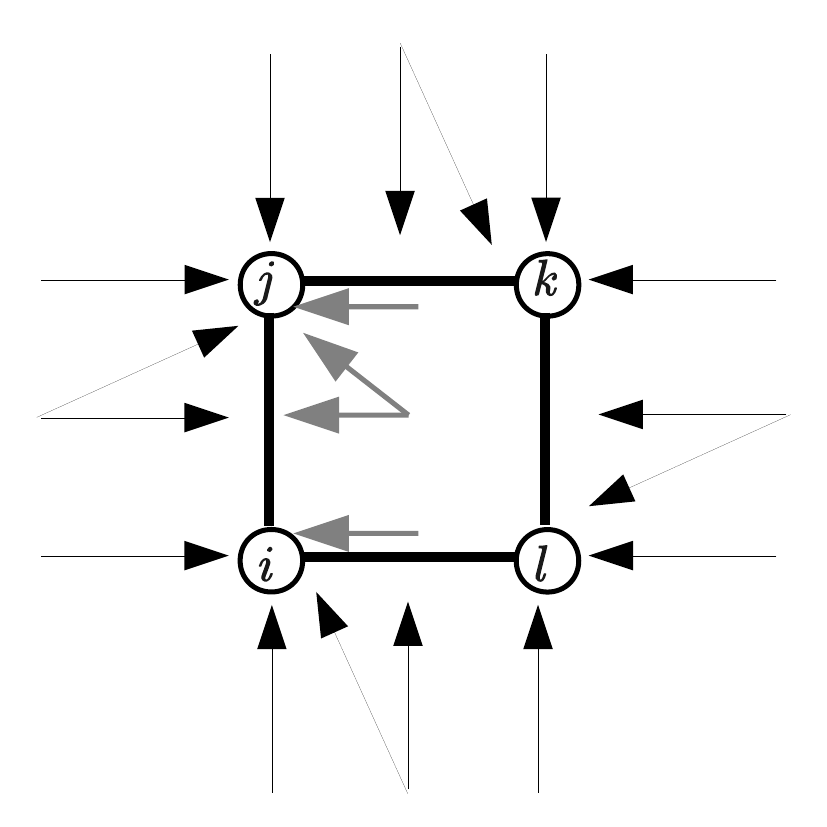}
\caption{Population dynamics basic step. Four 4-fields $(u_{L\to
i},U_{P \to ij},u_{P\to j},u_{L\to j})$ are taken at random from the
population, and new 4-fields are computed inside the square with random
couplings. The new 4-fields (one of which is shown in gray) are added
back to the population.}
\label{fig:gauge_mp}
\end{figure}

In the population dynamics method we have to evolve a population of 4-fields,
corresponding to two small-$u$ messages and a triplet $(U,u_1,u_2)$ message
arriving on the same pair of spins (see Fig.~\ref{fig:gauge_mp}).
Thanks to a local gauge symmetry, that
is worth breaking in order to improve the algorithm convergence properties
\cite{GBP_GF}, we can always set to zero one of the small-$u$ messages in
triplet (hence the name 4-field). In the average case, the correlation
between the Plaquette-to-Link and the Link-to-Spin fields is accounted
in the 4-fields structure, but different 4-fields are considered
uncorrelated around the plaquette. By randomly sampling the population and
the couplings distribution new 4-fields are computed as schematically
represented in Fig.~\ref{fig:gauge_mp}. After many iterations the population
stabilizes. The critical temperature is defined as the point where non-zero small $u$ messages
appear in the population of 4-fields and turns out to be very close to the
the value of $T_c$ computed in single instances. In ref. \cite{GBP_GF} 
these facts were reported as an interesting coincidence that now we extend 
to other values of $\rho$.

In the right panel of Figure \ref{fig:singleins} the upper dotted line marks the critical temperatures obtained by the gauge fixed (GF) population dynamics method. The key observation is that this temperature is quite close to the single instance critical temperature $T_c$ of the GBP message passing algorithm.  
Moreover in \cite{GBP_GF} we showed that the small discrepancy between the GF-SG critical temperature and the $T_c$ measured on single samples decreases by increasing the sample size. The closeness of these two temperatures suggests that the messages (4-fields) arriving on a plaquette in a 2D lattice are almost uncorrelated and thus lead to results similar to those obtained by a population dynamics, where messages are uncorrelated by construction.  So the critical temperature $T_c$ for a given large sample can very well be estimated from the average case GF population dynamics. At the same time, it suggests that fixing the gauge and keeping the correlation among the 4 fields in a 4-field message is important to get the right critical temperature, but in the average case replica calculation we can not fix the gauge and the correlation among the 4 fields is disregarded, since the distributions $Q(U,u_1,u_2)$ and $q(u)$ are independent. This is a weakness of the replica calculation in describing the actual behavior of message passing algorithm on given samples.

In the Bethe approximation, a population dynamics of Link to Spin fields reproduces exactly the same critical temperature found by the replica method \cite{MP1}. We tried to implement a new population dynamics, where all messages in a plaquette are updated at the same time, given the messages entering the plaquette, but the critical temperatures found do not compare well with BP--$T_c$. We also got not better results by simulating in Bethe approximation a population of the 2-fields $(u_1,u_2)$ that enter the plaquette from one side.

These facts point in either of two directions. The first possibility is that the closeness of the GF population dynamics critical temperature (GF-SG in Fig. \ref{fig:gauge_mp}) to the critical temperature $T_c$ in single instances is completely casual. The second is that the population dynamics is actually related to the single instance behavior. In this latter case, the fact that in Bethe approximation the population dynamics is useless in identifying $T_c$ implies that not only the correlation kept in the 4-fields is crucial, but also the presence of the $U$-fields, somehow overruling the actual interactions in the plaquettes, is very important.

\section{Non-linear regime}
\label{sec:ResultsLT}

Supported by the positive results of the previous sections, we look for the solution of the equations (\ref{eq:RSmess}) in the non-linear regime, below $T_\text{CVM}$. Still, the complete deconvolution of the second equation is beyond our technical capabilities and we reduce again the problem to that of computing the different moments of the functions $q$ and $Q$. However, now we keep the effect of the small messages beyond the linear regime. We show results for $\rho=0.5$ such that $m$ and $M_i(U)$ are zero. But the extension to more general cases is straightforward.

We start parametrizing $Q(U,u_1,u_2)$ in the following way:
\begin{equation}
Q(U,u_1,u_2) = a_0(U) \phi(u_1,u_2)\;,
\label{eq:parQ}
\end{equation}
where 
\begin{multline}
\phi(u_1,u_2) = (1 - p_U- q_U) \delta(u_1) \delta(u_2)
+ p_U \Big[ \delta(u_1-\sqrt{a}) \delta(u_2-\sqrt{a}) +\delta(u_1+\sqrt{a}) \delta(u_2+\sqrt{a})\Big] + \\
+ q_U \Big[ \delta(u_1-\sqrt{a}) \delta(u_2+\sqrt{a}) +\delta(u_1+\sqrt{a}) \delta(u_2-\sqrt{a})\Big]\;.
\label{eq:phia}
\end{multline}

\begin{figure}[htb]
\includegraphics[angle=0,width=0.3\textwidth]{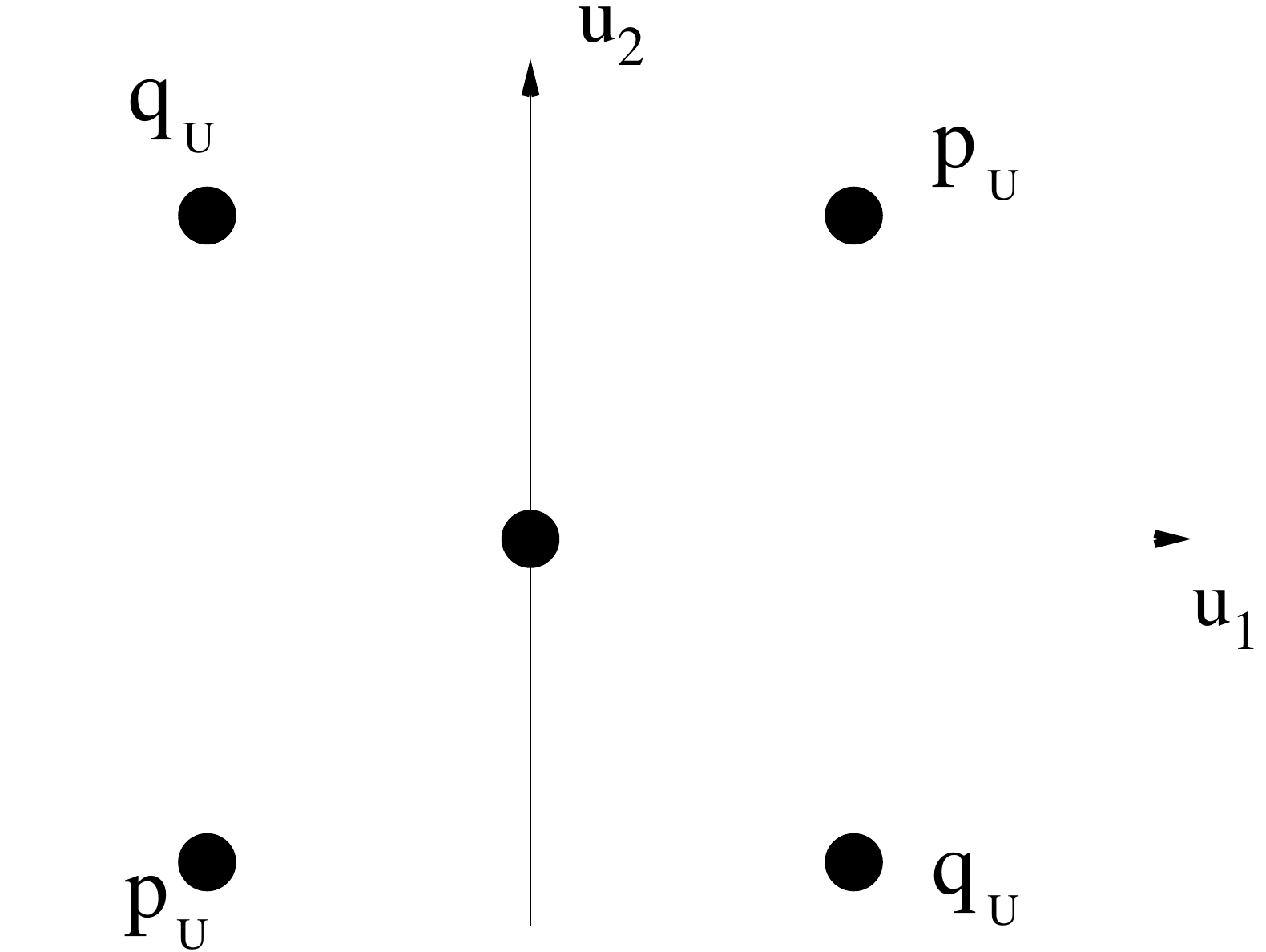}
\caption[0]{Schematic representation of the parametrization for $Q(U,u_1,u_2)$}
\label{fig:paramQ}
\end{figure}

This parametrization is sketched in Figure \ref{fig:paramQ}. It is important to point out that the function $\phi$ is not necessarily positive, and that the parameters $p_U$ and $q_U$ depend on $U$, so are functions themselves. We proceed writing these parameters in terms of the moments of the distribution $Q$. This is easily done substituting (\ref{eq:parQ}) in (\ref{eq:moments})
\begin{eqnarray}
a_{11} (U) &=& \int du_1 du_2 Q(U,u_1,u_2) u_1^2 = a_0(U) [ 2 a (p_U + q_U)]\;, \\
a_{12} (U) &=& \int du_1 du_2 Q(U,u_1,u_2) u_1 u_2 = a_0(U) [ 2 a (p_U - q_U)]\;.
\end{eqnarray}
such that
\begin{eqnarray}
p_U &=& \frac{a_{11}(U) + a_{12}(U)}{4\,a\,a_0(U)}\;,\\
q_U &=& \frac{a_{11}(U) - a_{12}(U)}{4\,a\,a_0(U)}\;.
\end{eqnarray}

\begin{figure}[htb]
\caption[0]{Schematic representation of the parametrization for $q(u)$}
\label{fig:paramq}
\end{figure}

Within this parametrization $a$ fixes the deviation from the paramagnetic solution of the distribution $Q$. But $a$ is  defined by the distribution of the small messages $q(u)$. Therefore we keep the consistency in the equations, without loosing physical insight, parametrizing also the $q(u)$ in terms of $a$. The simplest parametrization is sketched in Figure \ref{fig:paramq}. It reads:
\begin{eqnarray*}
q(u) = \frac{1}{2} \Big[ \delta(u -\sqrt{a}) + \delta(u -\sqrt{a}) \Big]&& \quad\text{if $a > 0$} \\
q(u) = 2 \delta(u) - \frac{1}{2} \Big[ \delta(u -\sqrt{|a|}) + \delta(u -\sqrt{|a|}) \Big]&& \quad\text{if $a < 0$}
\end{eqnarray*}

Note that the case $a<0$ must be taken into consideration because, since $Q$ is not necessarily positive defined \cite{tommaso_CVM}, during the message passing procedure $a$ may become negative. Now, specializing the computations to the case of the triangular lattice, the integrals over $R(U,u_1,u_2)$ in (\ref{eq:Maa}) take the form
\begin{multline}
a_{11}(U) = \int du_1 du_2 R(U,u_1,u_2) u_1^2 -a a_0(U)= \int dU^a dU^b a_0(U^a) a_0(U^b) \\\int d\vec{u}^ad\vec{u}^b \phi(u^a_1,u^a_2) \phi(u^b_1,u^b_2) \prod_{i=1}^4 q(u_i) \hat{u}_1^2 \delta(U - \hat{U}(\#))- a a_0(U)\;,
\label{eq:a11solved} 
\end{multline}
\begin{multline}
a_{12}(U) = \int du_1 du_2 R(U,u_1,u_2) u_1 u_2= \int dU^a dU^b a_0(U^a)  a_0(U^b) \\\int d\vec{u}^ad\vec{u}^b \phi(u^a_1,u^a_2) \phi(u^b_1,u^b_2) \prod_{i=1}^4 q(u_i) \hat{u}_1 \hat{u}_2 \delta(U - \hat{U}(\#))\;,
\label{eq:a12solved} 
\end{multline}
and $a$ satisfies
\begin{equation}
a = \int dU^a dU^b a_0(U^a)  a_0(U^b) \int d\vec{u}^ad\vec{u}^b \phi(u^a_1,u^a_2) \phi(u^b_1,u^b_2) \prod_{i=1}^5 q(u_i) \hat{u}^2(\#)\;,
\label{eq:asolved}
\end{equation}
where the integrals over $U$ are done using a standard population dynamics and the integrals over $\vec{u}$ can be computed exactly thanks to the previous ansatz [keep in mind that $\phi(u^a_1,u^a_2)$ is given by (\ref{eq:phia})].
The analysis for any other lattice is completely equivalent. Independently of the structure of the plaquettes, or the lattice dimensions the previous ansatz is always valid and the fixed point equations can be always reduced to expressions similar to (\ref{eq:a11solved})-(\ref{eq:asolved}). Only the computational effort may change. For example, while in equations (\ref{eq:a11solved}) and (\ref{eq:a12solved}) we integrate over two $U$ messages, $U^a$ and $U^b$, in the square lattice we will need a third message to integrate over. However, from the results obtained in the previous section  we do not expect any gain in physical insight from studying  the square lattice and we concentrate our efforts on the triangular lattice.

\begin{figure}[htb]
\includegraphics[angle=0,width=0.55\textwidth]{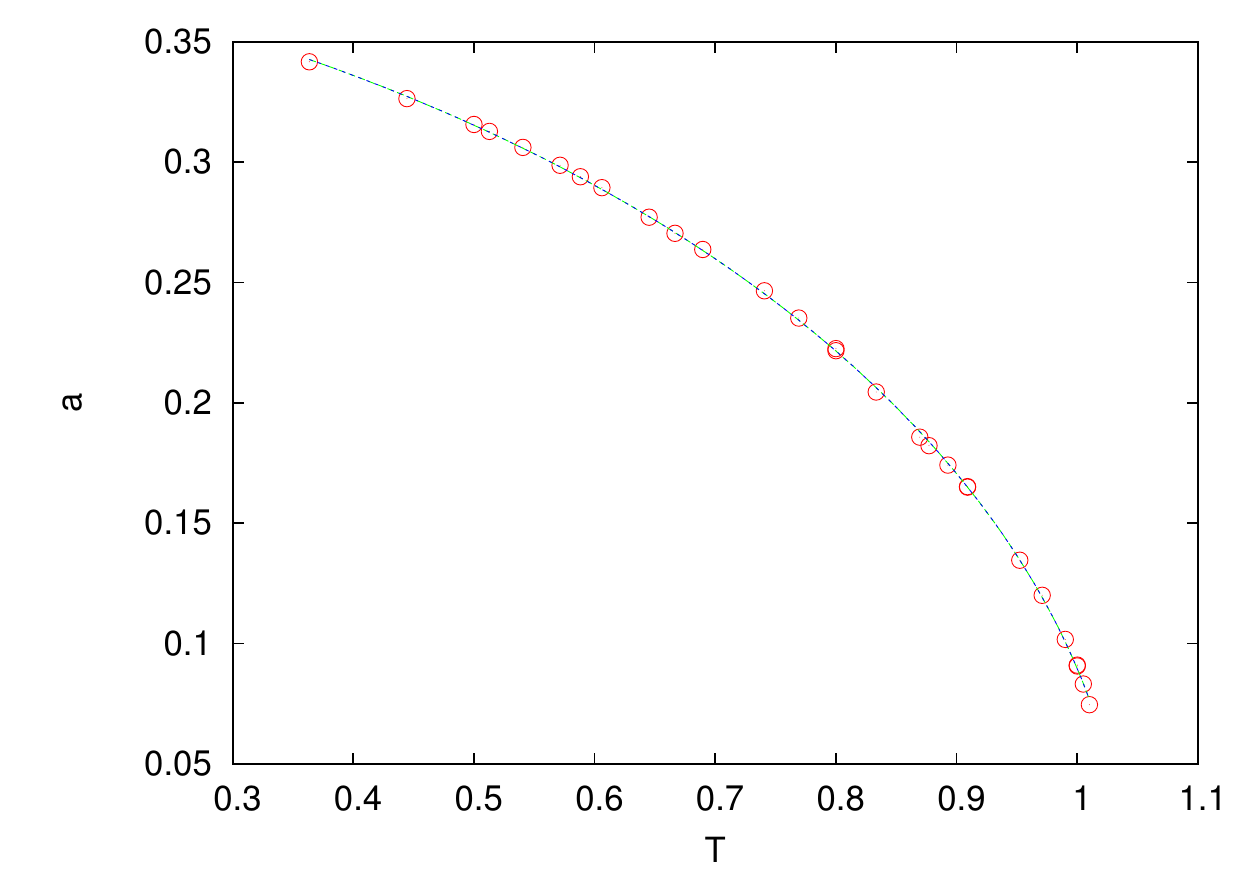}
\caption[0]{The $a$ parameter for the triangular lattice in the low temperature phase. The curve is a fit behaving as $(T_\text{CVM}-T)^{1/2}$ close to the critical point.}
\label{fig:avsbeta}
\end{figure}

Our first result is presented in Figure \ref{fig:avsbeta} where we present the dependence of $a$ with $T$ below $T_\text{CVM}$. Note that the data is compatible with a behaviour of the form $a \propto (T_\text{CVM} -T)^{1/2}$, although analytical arguments would suggest a linear behavior in $(T_\text{CVM}-T)$, much as in the Bethe approximation case. It may be that the linear coefficient is actually very large but we did not further investigate this point because it would require a consistent increase of numerical precision in the critical region.

In presence of an external field, the symmetry which allow for the existence of polynomial algorithm to solve the 2D EA model \cite{Pfafian,middleton09} breaks down. On the other hand the method based of the replica CVM equations can be perfectly used also in presence of an external field: the equations remain practically the same, with the only difference that the external field must be added to the local field $h$ in all the expressions above. We leave for the interested reader to prove this. 

\begin{figure}[htb]
\includegraphics[angle=0,width=0.45\textwidth]{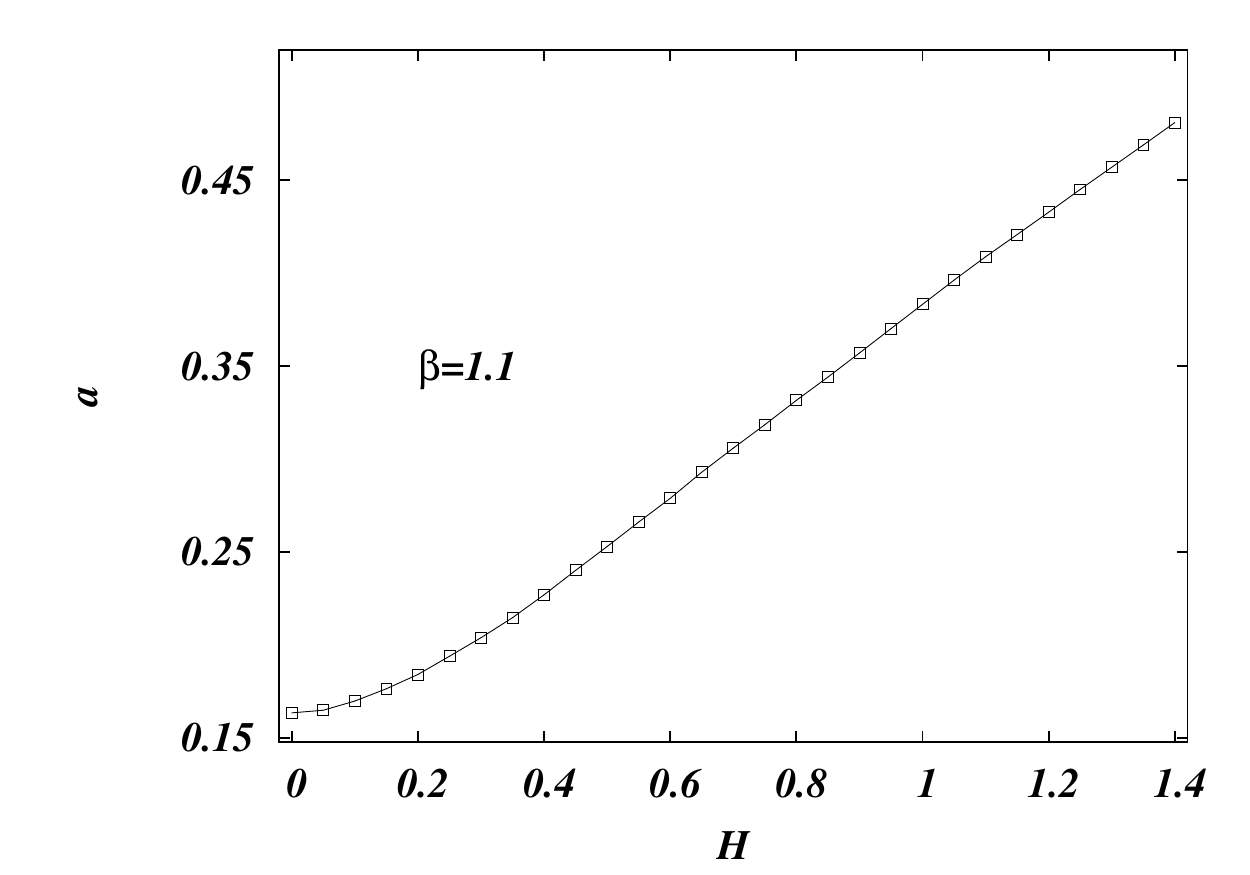}
\includegraphics[angle=0,width=0.45\textwidth]{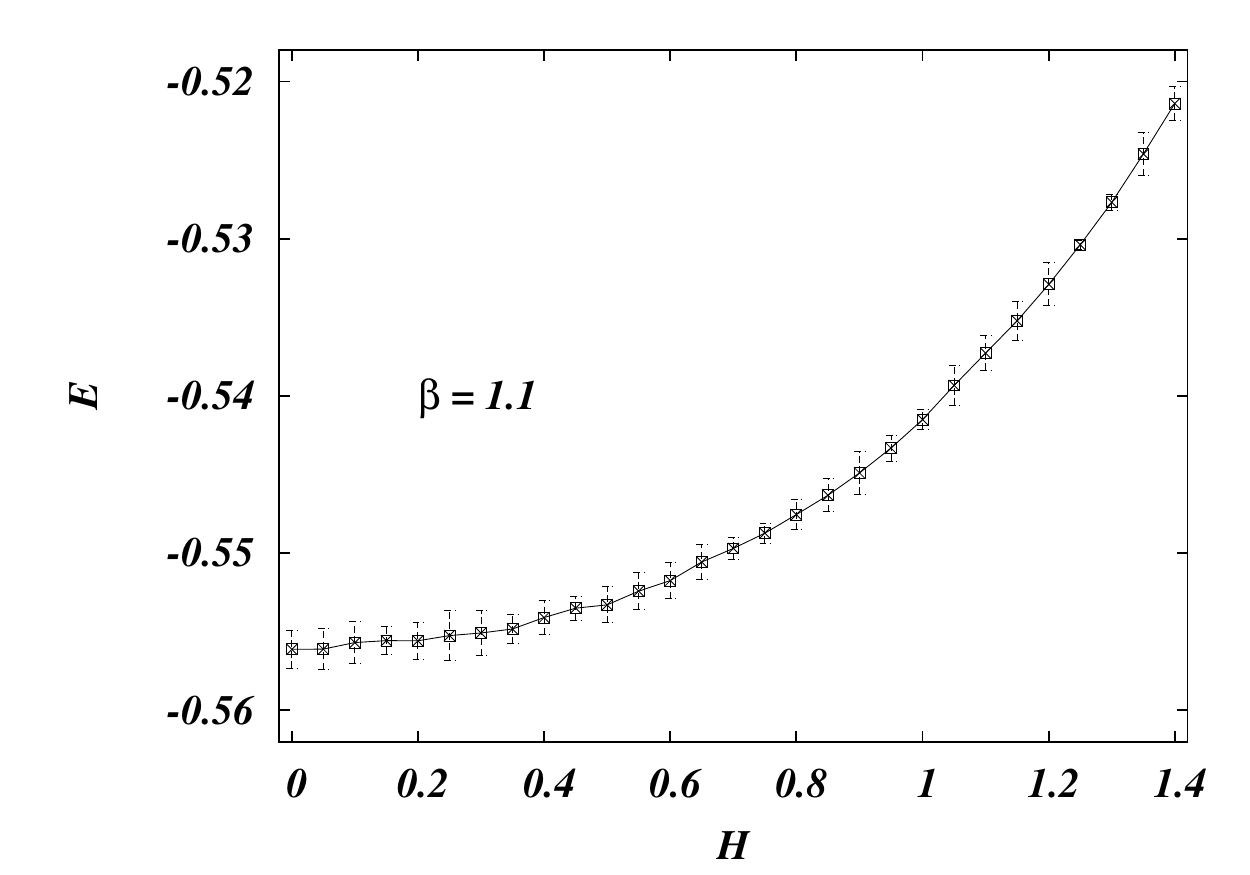}
\caption[0]{The $a$ parameter (left) and the energy $E$ (right) versus the external field $H$. The temperature is slightly below the Paramagnetic-Spin Glass phase transition ($\beta=1.1$).}
\label{fig:aEvsH}
\end{figure}

We study the model in the presence of an external magnetic field, near, but below, the transition temperature and show (see Figure \ref{fig:aEvsH}) that both $a$ and the energy $E$ go as $H^2$ close to the transition. Our results, although approximate, can be considered as a good starting point to study the role of the external field in finite dimensional lattices.

\begin{figure}[htb]
\includegraphics[angle=0,width=0.45\textwidth]{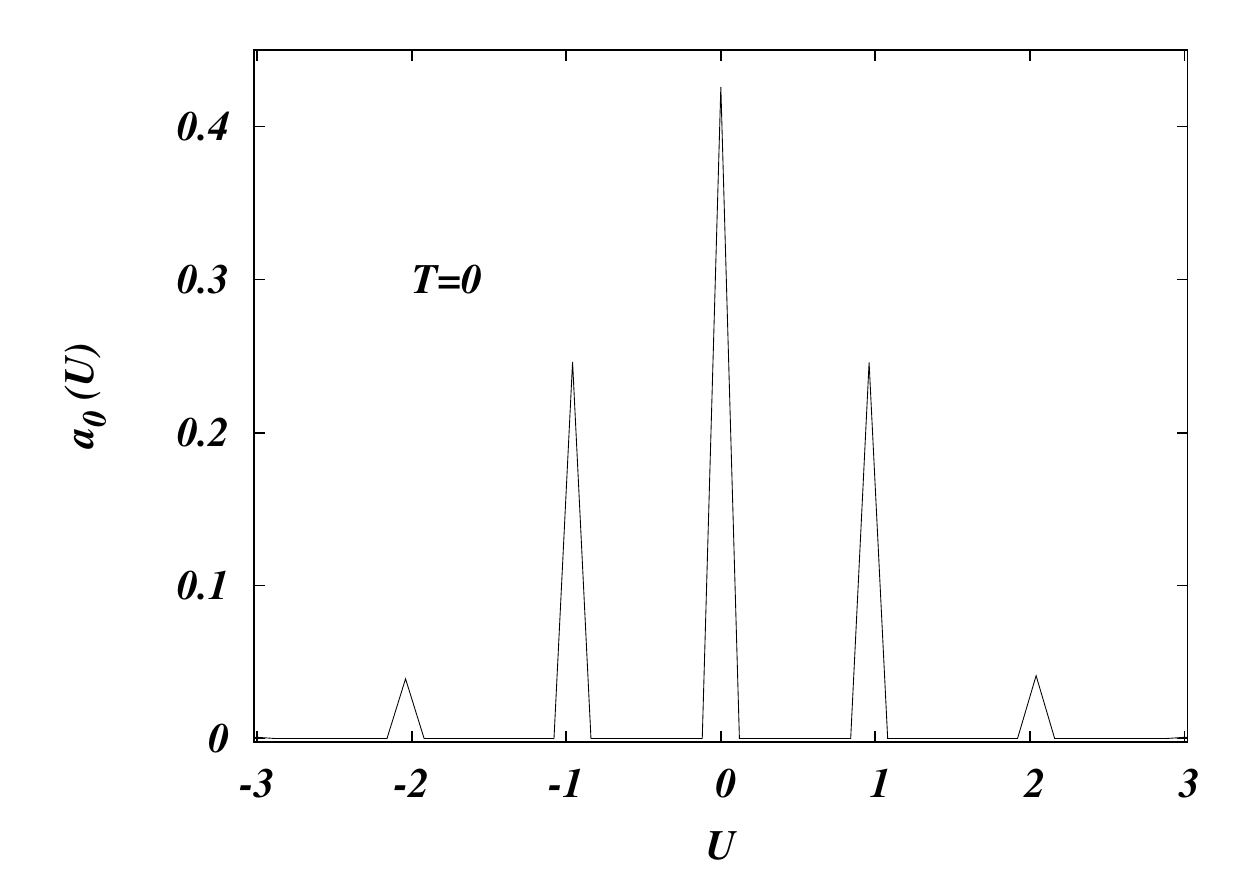}
\includegraphics[angle=0,width=0.45\textwidth]{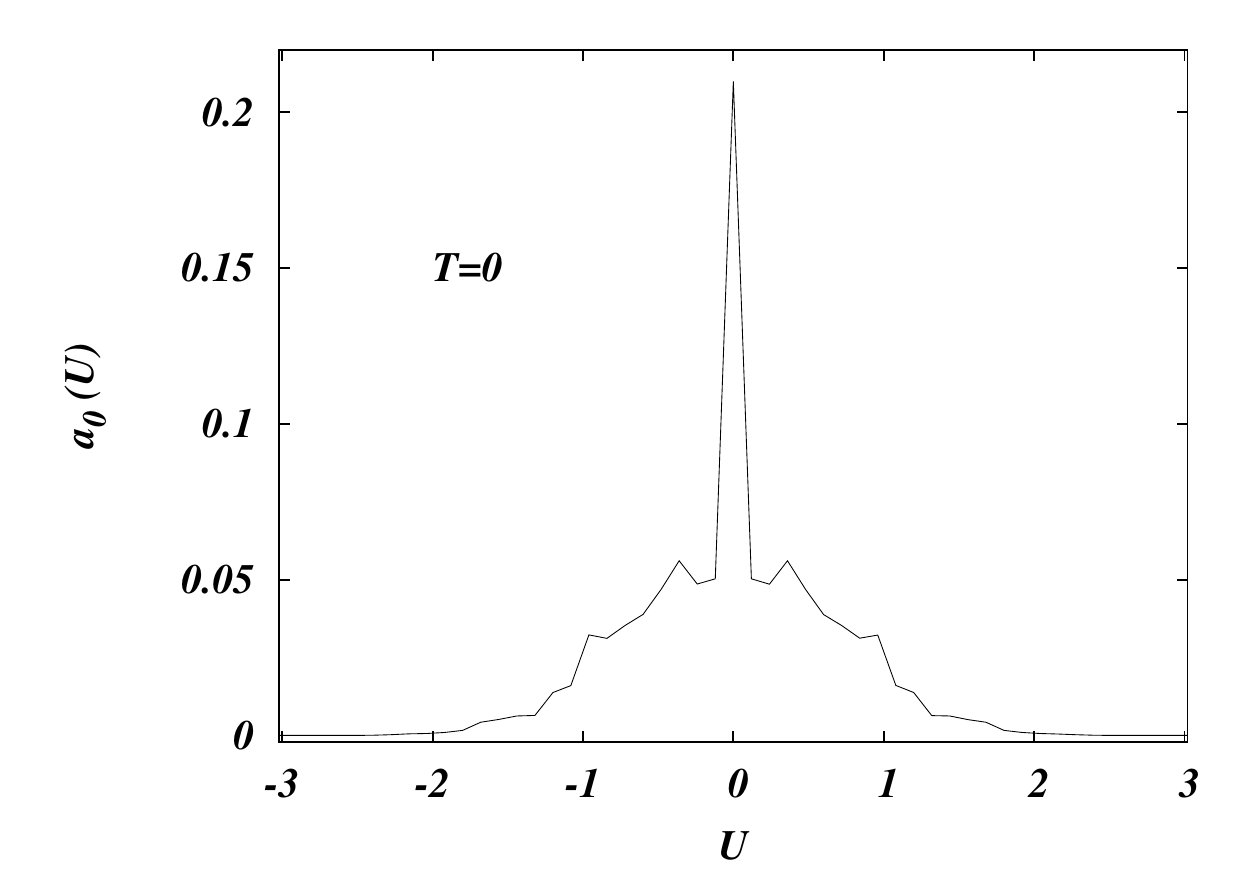}
\caption[0]{Distribution $a_0(U)$ of the messages $U$ at zero temperature, using the paramagnetic ansatz [left] and the non-linear ansatz of eq.(\ref{eq:parQ}) [right].}
\label{fig:a0}
\end{figure}

Finally we prove that this technique can be extended to zero temperature and provides non trivial informations also in that limit. In Figure \ref{fig:a0} we show the structure of $a_0(U)$ considering a paramagnetic ansatz (left) where $\phi(u_1,u_2)=\delta(u_1)\delta(u_2)$ and $q(u)=\delta(u)$, and after reaching the fixed point of equations (\ref{eq:a11solved})-(\ref{eq:asolved}) (right). The paramagnetic solution has a structure very similar to the one found in the study of the EA model on a Bethe lattice \cite{MP2}. This is not surprising since within the paramagnetic ansatz the problem is equivalent to a Bethe approximation on the dual lattice (see our previous work \cite{dual} for a larger discussion on this subject). On the other hand, the structure of $a_0(U)$ when non-linear effects are considered is richer. While the $U=0$ peak still dominates the distribution, and there is some reminiscence of other peaks, now the distribution spreads over non-integer values. This is probably one of the more remarkable mathematical consequences of the Kikuchi approximation. It is enough to consider the equation for $\hat{U}$ in the presence of small $u$'s, to understand that it is not possible to keep the self-consistency of the equations with distributions supported in the integers (even at $T=0$). This unavoidable fact make the computations at $T=0$ as heavier as the computations at finite temperature and further contributes to make the Kikuchi approximation harder to deal than the Bethe approximation.

\begin{figure}[htb]
\includegraphics[angle=0,width=0.45\textwidth]{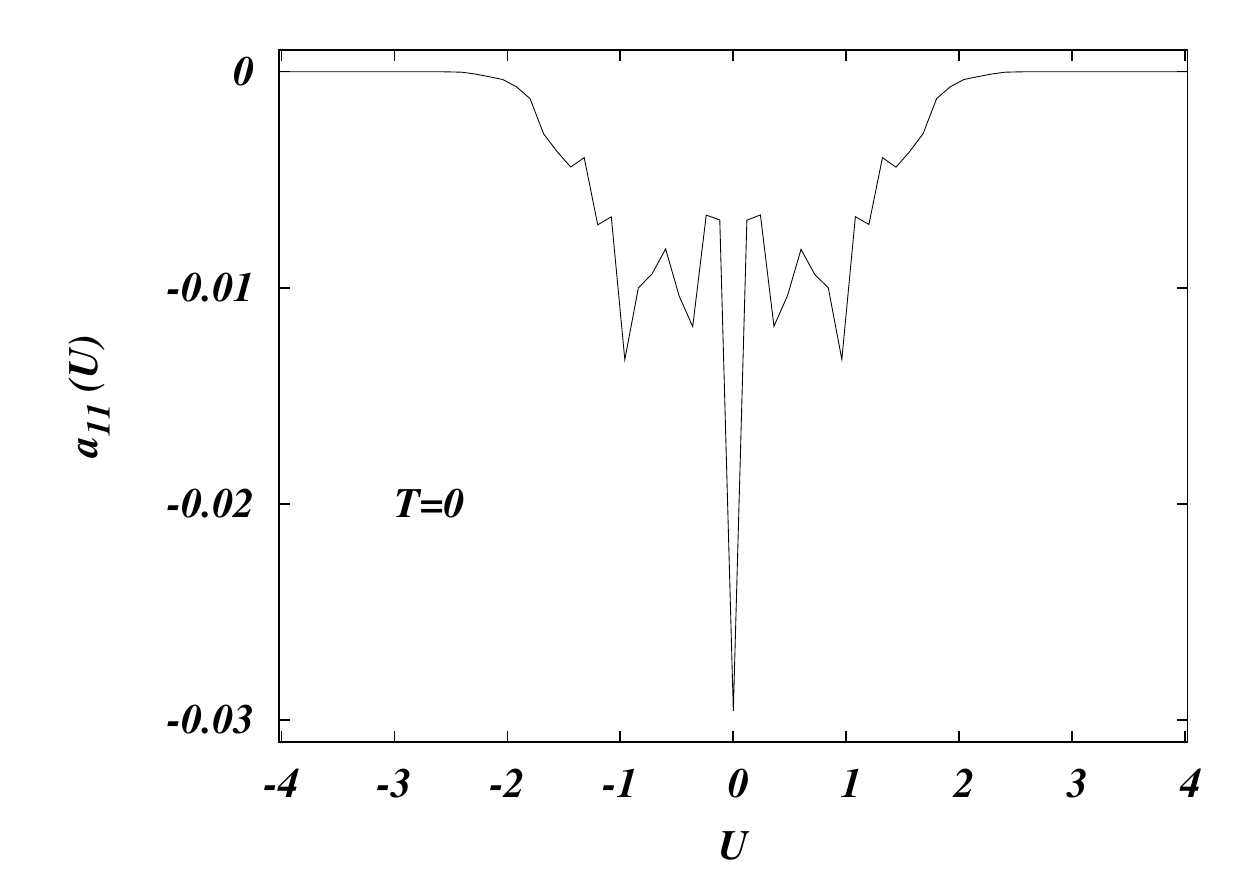}
\includegraphics[angle=0,width=0.45\textwidth]{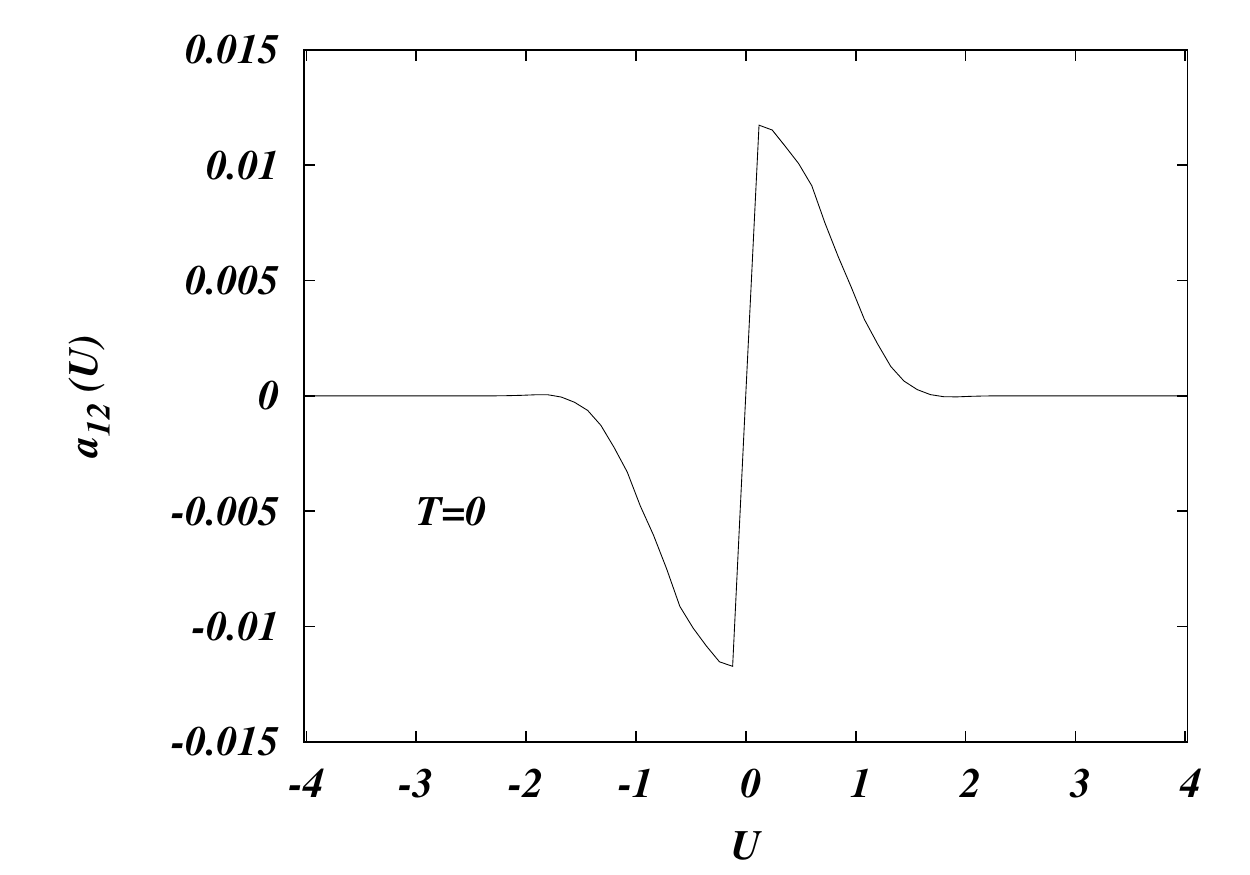}
\caption[0]{Self-correlations $a_{11}(U)$ (left) and cross-correlations $a_{12}(U)$ (right) of small $u$ messages, using the non-linear ansatz in eq.(\ref{eq:parQ})}
\label{fig:a11a12}
\end{figure}

To further explore the role of $\phi(u_1,u_2)$ we plot $a_{11}(U)$ (left) and $a_{12}(U)$ (right) in Figure \ref{fig:a11a12}. It is interesting to note that while $a_{11}(U)$ presents a structure with multiple peaks resembling the structure of $a_0(U)$, there is no clear evidence of such a structure in $a_{12}(U)$. This suggests that self-correlations of the small $u$ messages strongly depend on $a_0(U)$, much more than what the cross-correlations do (these show a smooth curve at every $U$).

\section{Conclusions}
\label{sec:Conc}

We study typical properties of the 2D Edwards-Anderson model with the Replica Cluster Variational Method at the RS level. Using a linearized version of the self-consistency equations we have obtained the $\rho$ vs $T$ phase diagram on the square and triangular lattices. We show that this phase diagram resembles much better the theoretical predictions, than the one obtained using the Bethe approximation: the SG critical temperature is lower, the tricritical point is closer to the exact value and the SG-Ferro phase boundary looks similar to theoretical expectations. Moreover, we present numerical evidences supporting the idea that the temperature below which the average case computation predicts the existence of a spin-glass phase ($T_\text{CVM}$) is also the temperature at which GBP algorithms stop converging.  We apply to the triangular lattice a method to solve the RS equations in the non-linear regime, i.e., at very low temperatures. The method does work and we show results at $T=0$ and in the presence of an external magnetic field. All these results suggest that the replica CVM can be used to study finite-dimensional spin glasses, and hopefully in higher dimensions ($D>2$) the approximation should provide an even better description of the low temperature phase.

\begin{acknowledgments}

F.R.-T. acknowledges the hospitality of LPTMS at Univerit\'e Paris Sud during the completion of the manuscript, and financial support by the Italian Research Minister through the FIRB Project No. RBFR086NN1 on ``Inference and Optimization in Complex Systems: From the Thermodynamics of Spin Glasses to Message Passing Algorithms''.

\end{acknowledgments}

\section*{Appendix: Triangular lattice}
\label{app:trian}

We report here the expressions for the first and second moments of $Q(U,u_1,u_2)$ in the case of the triangular lattice.
\begin{multline}
M_1(U) = \langle\int dU^a dU^b \delta\big(U - \arctanh[\tanh(\b(J^a+U^a))\tanh(\b(J^b+U^b))]/\b\big)\\
\Bigg[ \frac{M_1(U^a) a_0(U^b) + a_0(U^a) M_1(U^b)}{2} + 2 m \tanh(\b J_+) +
\tanh(\b J_+) \frac{M_1(U^a) + M_1(U^b)}{2} \Bigg] \rangle_J\;,
\label{eq:M1trian}
\end{multline}
\begin{multline}
a_{11}(U) = \langle\int dU^a dU^b \delta\big(U - \arctanh[\tanh(\b(J^a+U^a))\tanh(\b(J^b+U^b))]/\b\big)\\
\Bigg[ \frac{a_{11}(U^a) a_0(U^b) + a_0(U^a) a_{11}(U^b)}{2} + 
2 m \tanh(\b J_+) [M_1(U^a) + M_1(U^b)] + \\
+ 2 m \tanh(\b J_-) [M_1(U^a) - M_1(U^b)] +
\tanh(\b J_+) M_1(U^a) M_1(U^b) +\\ 
\frac{1}{2}\tanh(\b J_+) [a_{12}(U^a) + a_{12}(U^b)] + \frac{1}{2}\tanh(\b J_-) [a_{12}(U^a)-a_{12}(U^b)]+\\
(a + 3 m^2) [\tanh^2(\b J_+) + \tanh^2(\b J_-) ]+
2 m [\tanh^2(\b J_+) + \tanh^2(\b J_-) ][M_1(U^a) + M_1(U^b)]+\\
\frac{1}{4} [\tanh^2(\b J_+) + \tanh^2(\b J_-)] [a_{11}(U^a) + a_{11}(U^b) + 2 M_1(U^a) M_1(U^b)]
\Bigg]\rangle_J\;,
\label{eq:a11trian}
\end{multline}
\begin{multline}
a_{12}(U) = \langle\int dU^a dU^b \delta\big(U - \arctanh[\tanh(\b(J^a+U^a))\tanh(\b(J^b+U^b))]/\b\big)\\
\Bigg[ M_1(U^a) M_1(U^b) + 
2 m [\tanh(\b J_+)(M_1(U^b)+M_1(U^a)) + \tanh(\b J_-)(M_1(U^b)-M_1(U^a)) ]+\\
\frac{1}{2}\tanh(\b J_+) [2 M_1(U^a) M_1(U^b)+a_{12}(U^a)+a_{12}(U^b)] +\tanh(\b J_-) \frac{a_{12}(U^b) - a_{12}(U^a)}{2} +\\
(a + 3 m^2) (\tanh^2(\b J_+)-\tanh^2(\b J_-)) + 2 m [ \tanh^2(\b J_+)-\tanh^2(\b J_-) ] [M_1(U^a)+ M_1(U^b)]+\\
\frac{1}{4} [\tanh^2(\b J_+)-\tanh^2(\b J_-) ] [a_{11}(U^a) + a_{11}(U^b) + 2 M_1(U^a) M_1(U^b) ]
\Bigg]\rangle_J\;,
\label{eq:a12trian}
\end{multline}
where $J_+ = (J^a + U^a) + (J^b+ U^b)$,  $J_{-} = (J^a + U^a) - (J^b + U^b)$.

\bibliography{bibliografia}
\end{document}